%% file: main.tex
\documentclass[conference]{IEEEtran}
\usepackage{xcolor}
\usepackage{hyperref}
\usepackage{graphicx}
\usepackage[normalem]{ulem}
\usepackage{algorithm}
\usepackage{algorithmic}
\usepackage{stmaryrd}
\usepackage{tikz}
\usepackage{amsfonts}
\usepackage{amsmath}
\usepackage{mathtools}

\ifCLASSINFOpdf
\else
\fi
\begin{document}
%
\title{Rotating labeling of entropy coders for synthetic DNA data storage}

\author{\IEEEauthorblockN{Xavier Pic, Eva Gil San Antonio, Melpomeni Dimopoulou, Marc Antonini}
\IEEEauthorblockA{\textit{I3S laboratory, Côte d’Azur University and CNRS}\\
UMR 7271\\Sophia Antipolis, France\\
xpic@i3s.unice.fr}}

%


\maketitle



%
\IEEEpeerreviewmaketitle
\input{abstract}
\vspace{-\baselineskip}
\input{introduction}
\vspace{-\baselineskip}
\input{context}
\vspace{-0.5\baselineskip}
\input{coding}
\vspace{-0.5\baselineskip}
\input{performance_results}
\input{stats_review}
\vspace{-\baselineskip}
\input{conclusion}
\vspace{-\baselineskip}

\bibliographystyle{IEEEtran}
\bibliography{refs}
%
%
%

\end{document}

%% file: abstract.tex
\begin{abstract}
Over the past years, the ever-growing trend on data storage demand, 
has motivated research for alternative systems of data storage.
Because of its biochemical characteristics, synthetic DNA molecules are 
considered as potential
candidates for a new 
storage paradigm.
Because of this trend, 
several coding solutions have been proposed over the past years for the storage of digital information into DNA.
Despite being a 
promising 
solution, DNA storage faces two major obstacles: the large cost of synthesis and the noise introduced during sequencing. Additionally, this noise increases when biochemically defined coding constraints are not respected: avoiding homopolymers and patterns, as well as balancing the GC content.
This paper describes a novel entropy coder which can be embedded to any block-based image-coding schema and aims to robustify the decoded results. Our proposed solution introduces variability in the generated quaternary streams, reduces the amount of homopolymers and repeated patterns to reduce the probability of errors occurring. 
While constraining the code to better satisfy the constraints would degrade the compression efficiency, in this work, we propose an alternative method to further robustify an already-existing code without affecting the compression rate.
To this end, we integrate the proposed entropy coder into 
four existing JPEG-inspired DNA coders.
We then evaluate the quality ---in terms of biochemical constraints--- of the encoded data for all the different methods by providing specific evaluation metrics.
\end{abstract}

%% file: introduction.tex
\section{Introduction}
\label{sec:intro}
Images represent a big percentage of the data stored in data centers and a large majority of these images is considered as ``cold" (very infrequently accessed). For that reason, it is crucial to develop image coders that have a good compression performance and are adapted to the problematics inherent to DNA data storage. Over the past years, some coders \cite{Church, Goldman2013, Aeon}, were developed to encode data into DNA. An international JPEG standardization group, JPEG DNA, was formed in 2020 to address the problem of standardizing the coding of still images on molecular media, with a standard expected in 2025.
In this paper, we present a novel block-based coding solution
for DNA data storage. We then integrate it in different image codecs for synthetic DNA data storage, all inspired by the legacy JPEG 
algorithm.

In section \ref{sec:context} we introduce some concepts around DNA data storage as well as the existing JPEG-inspired codecs proposed to 
address this new field of research. Section \ref{sec:coding} explains the interest of further robustifying entropy coders against sequencing errors. The proposed block-based coding scheme is then explained while proving it does not deteriorate the compression of the original method. The coding scheme is then integrated into some existing JPEG-based image codecs presented in section \ref{sec:context}.
Finally, in section \ref{sec:CompressionPerformances} and \ref{sec:quality} the paper shows performance results. We experimentally show that the modified codecs don't loose any performance in compression, and statistically analyse the quality of the encoded data with regards to the 
DNA
coding constraints. The modified codecs have a better GC-content than the original ones and the long homopolymers have been removed from the generated oligos.

%% file: context.tex
\section{Context}
\label{sec:context}
\subsection{DNA data storage}
Today, the digital world relies on increasingly large amounts of data, stored over periods of time ranging from a few years to several centuries. 
With current storage media reaching its density limit and the exponential growth of digital information, the search for a new storage paradigm has become of utmost importance.
One of the most promising candidates up to this date is to store information in the form of DNA molecules, which provide very dense
storage (1 $EB/mm^3$) that is also stable over long periods if stored under the right conditions
\cite{DNARAM}.

The process of DNA data storage starts by encoding data into a quaternary stream composed by the four DNA symbols or nucleotides (A, C, T, and G). The encoded data is then physically synthesised into DNA 
strands (oligos)
that are then stored into a safe environment. When data has to be read back, the stored molecules are amplified so as to obtain many copies of each original sequence 
and the content is deciphered using DNA sequencers. This data is then decoded back into the form of the original binary file. However, the biochemical processes involved in the above end-to-end storage process introduce some coding constraints \cite{Church} that, if not respected, dramatically increase the probability of an error occurring (insertion, deletion and/or substitution). These constraints comprise avoiding homopolymers (i.e. use of the same symbol more than 3 consecutive times), repetition of patterns and unbalanced GC content.

Algorithms specifically designed for DNA data storage which respect the biochemical constraints generally show better reliability \cite{Goldman2013}. Although constrained coding helps reducing the apparition of errors, it does not ensure an error free coding. To tackle the problems of errors, an important number of researchers \cite{Welzel, Imperial, DNASmart} are currently studying the development of accurate error models for synthetic DNA data storage.
Furthermore, since the cost of DNA synthesis is relatively high, it is also important to take advantage of optimal compression, which can be achieved before synthesizing the sequence into DNA. 
To decrease the synthesis cost of image storage into DNA, several works have been using the so-called ``transcoding'' method which is based on classical compression protocols such as JPEG for reducing the image redundancies, then encoding each byte of the produced bitstream
into DNA words. However, it is clear that such a compression schema is sub-optimal as the compression is optimised with respect to a binary code 
and therefore it is not adapted to the quaternary nature of DNA.
To tackle the above issue, 
among other relevant works, in \cite{DNAcoding} the authors proposed a specific image coding 
algorithm, adapted to the needs of DNA data storage, that efficiently encodes images into a quaternary constrained code.

\subsection{JPEG-inspired coding methods}
\begin{figure}
    \centering
    \includegraphics[scale=0.09]{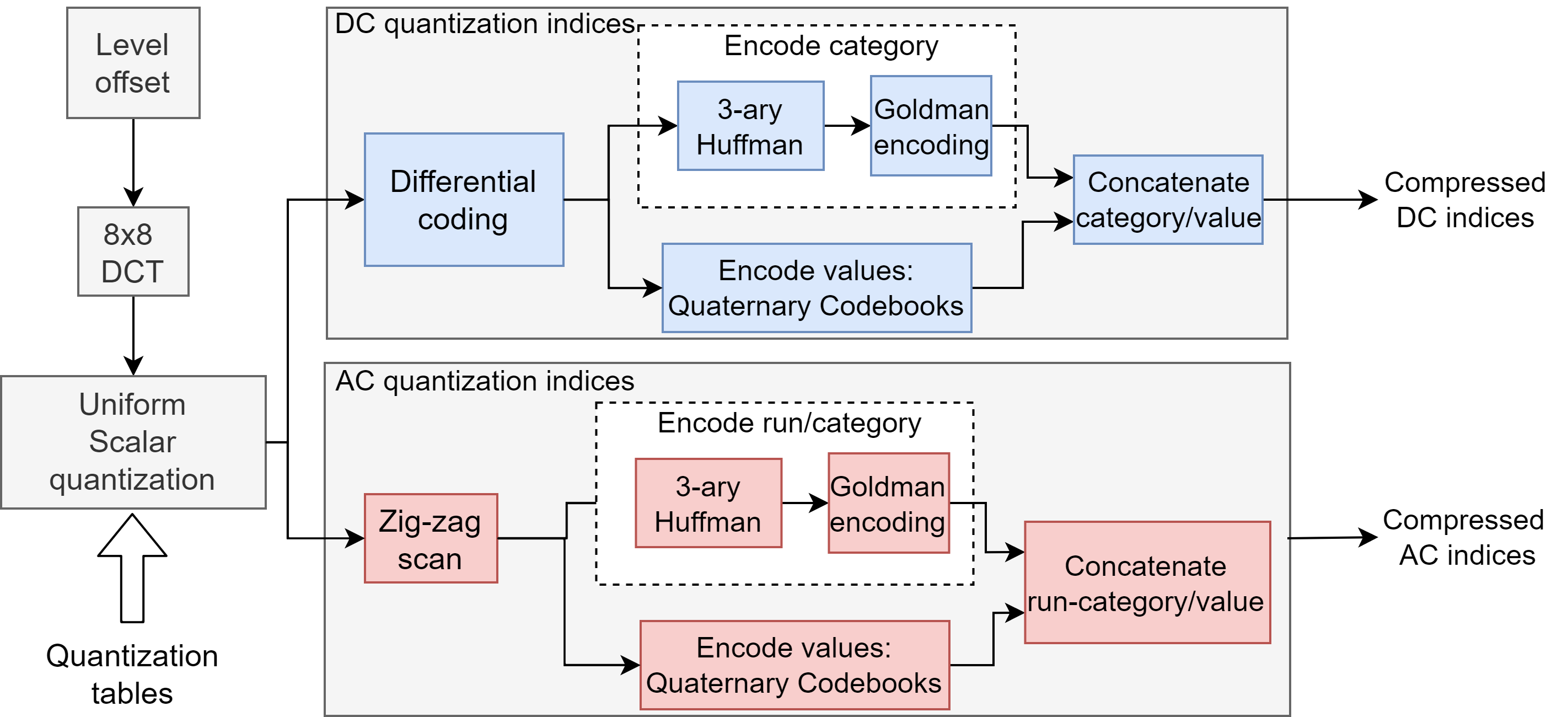}
\vspace{-\baselineskip}
    \caption{Workflow of a JPEG-inspired codec \cite{DNAcoding,CTC}} 
    \label{fig:workflow}
\vspace{-\baselineskip}
\end{figure}

All the JPEG-inspired coding methods described here are based on \cite{DNAcoding}, 
whose
workflow is described in Fig. \ref{fig:workflow}.
\subsubsection{Principle}
The main principle of these coders is to encode a sequence of DCT coefficients into DNA using a combination of an entropy and a fixed-length constrained coder.
The coders are constrained so as to be as compliant as possible with regards to the biochemical constraints.
The run/category is encoded with the entropy coder and the value with the fixed-length coder (see \cite{DNAcoding}).
\subsubsection{Entropy coders}
Entropy coding methods are variable-length lossless coding 
methods which encode the source data thanks to a code --- a set of codewords --- generated from a frequency table representing the number of appearances of every symbol in the source. The length of the codewords representing every symbol varies according to the probability of their appearance: frequently appearing symbols are associated to shorted codewords while the less frequent ones are assigned to longer codewords.
There are two possible entropy coders that can be used to encode the run/category 
of the coefficients: Huffman/Goldman \cite{Goldman2013} and the Shannon Fano based constrained coder we proposed in \cite{SFC4}. The latter is more performant in terms of compression rate.
\subsubsection{Source type}
When one encodes an image into DNA with a JPEG-inspired algorithm, there are two possibilities. If the input of the algorithm is a raw image, the algorithm computes the DCT coefficients and then encodes them \cite{DNAcoding} into DNA. This encoding paradigm is called source coding. Otherwise, the input of the coding algorithm can be an already compressed JPEG binary file. In this case, the bitstream is decoded to obtain the sequence of DCT coefficients that will then be encoded into DNA \cite{JPEGDNABCT}. We will refer to this paradigm as transcoding.


%% file: coding.tex
\section{Coding solution}
\label{sec:coding}
\subsection{Motivation}
\begin{figure*}[ht!]
    \centering
    \includegraphics[scale=0.65]{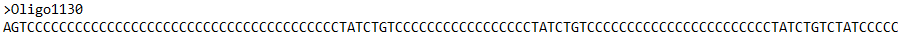}
\vspace{-0.5\baselineskip}
    \caption{Problematic encoding cases in JPEG-inspired codecs: example of homopolymers created at high compression rates.}
    \label{fig:JPEGDNAprob}
\vspace{-0.5\baselineskip}
\end{figure*}

\begin{figure}
    \centering
\vspace{-0.5\baselineskip}
    \includegraphics[scale=0.16]{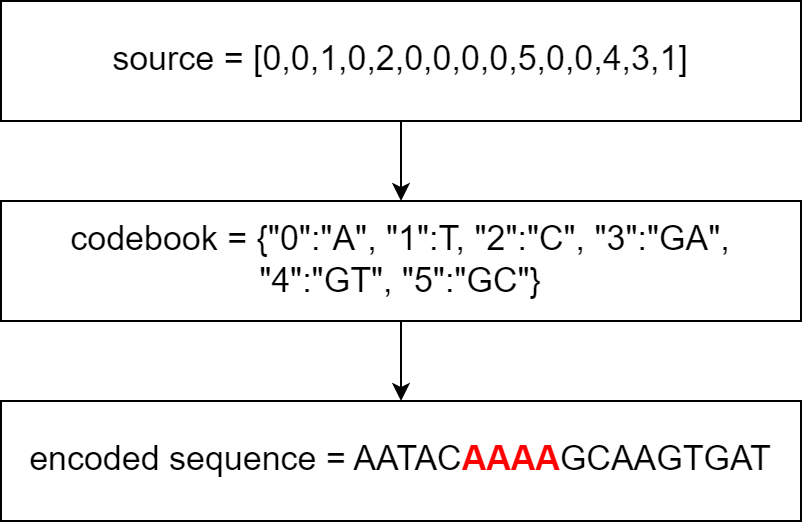}
\vspace{-0.5\baselineskip}
    \caption{Problematic case for an entropy coder. In red, a homopolymer caused by the repetition of a very frequent symbol in the source.}
    \label{fig:problem}
\vspace{-1\baselineskip}
\end{figure}

The principle of the novel coding solution is to avoid using the same code --- set of codewords --- consecutively to avoid creating patterns or homopolymers.
In the case where a symbol is very frequently found in the source, due to its high probability, the entropy coder will associate a short codeword to it, which in extreme cases can lead to a codeword consisting of one nucleotide. 
If this symbol is repeated consecutively in the source, it will create a homopolymer. Figure \ref{fig:problem} represents this case.
If the symbol is a bit less frequent, the entropy coder will associate a codeword of length two or three to this symbol, and when a repetition of this symbol will occur, a short pattern will be created, also violating
the defined biochemical constraints. A real example of these phenomenon can be found when encoding images into DNA with a JPEG-inspired algorithm at very high compression rates, like shown in Fig.\ref{fig:JPEGDNAprob}.

\subsection{Principle}
In order to avoid these homopolymers and repetitions of patterns, 
we propose to introduce variability in the generated quaternary streams by generating several equally peformant codes and alternating between them during encoding. 
Thus, since the symbols will not always be associated to the same codewords, we reduce the risk of creating homopolymers when very frequent symbols are heavily repeated in the source. 

\begin{figure}
    \centering
    \includegraphics[scale=0.16]{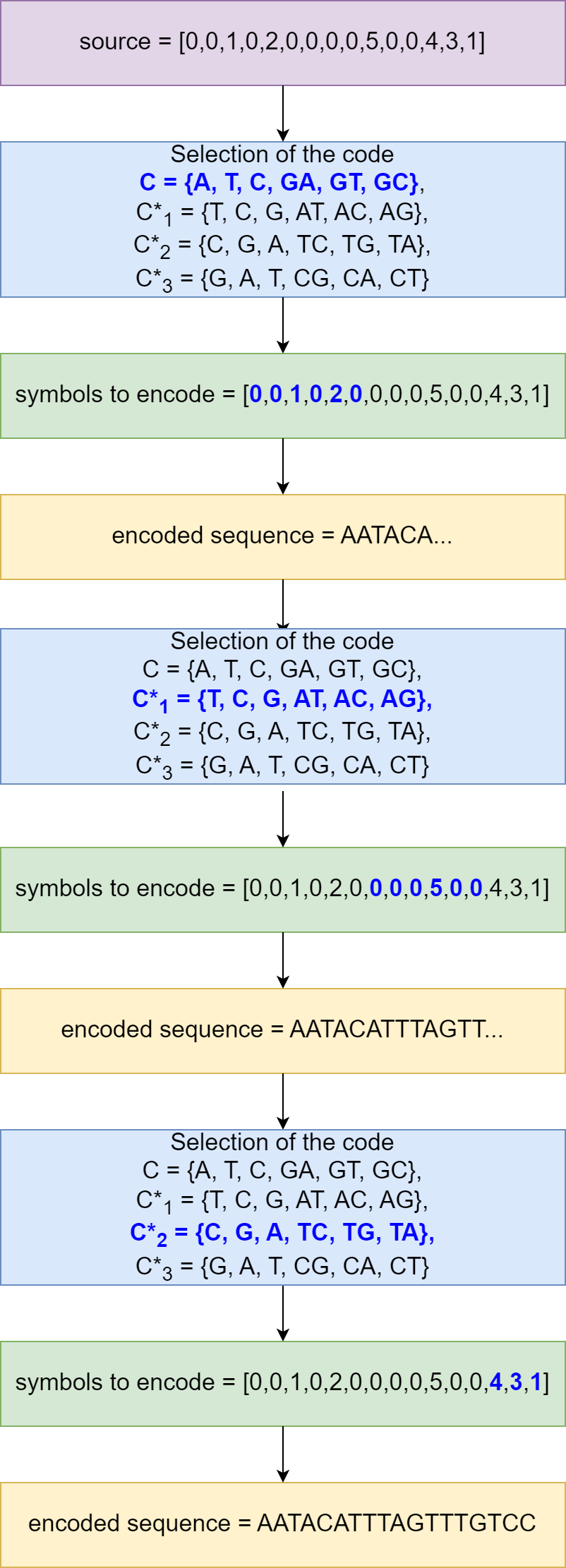}
\vspace{-0.5\baselineskip}
    \caption{Encoding example using rotating codes to avoid homopolymers. The code is changed every six symbols.
    }
    \label{fig:non_erroneous}
\vspace{-1\baselineskip}

\end{figure}

An entropic code $C$ is mapped to three associated entropic codes $C^*_1$, $C^*_2$ and $C^*_3$, with three mapping functions $M_1$, $M_2$ and $M_3$ described in Algorithms \ref{algo:cdbks_gen} and \ref{algo:cw_mod} and defined as: $\forall k \in [1, 3],$\\
\vspace{-1.5\baselineskip}
\begin{align*}
    M_k\colon & C \to C^*_k\\
    &c \mapsto c^*_k = switchLetters(c, k)
\vspace{-2\baselineskip}
\end{align*}
The code used to encode the symbols can be changed periodically. When the source is a simple sequence of symbols (i.e. non block-based), one can change the code after having encoded a fixed amount of symbols, for example every six symbols like the solution presented in Fig. \ref{fig:non_erroneous}.
\par
This novel encoding method relies on two important processes that will be further explained in the following sections: the generation of the different codes that can be used to encode the symbols, and the decision of which code to use. Both processes have to respect some prerequisites to make the whole encoding method a viable solution. 
First, the proposed coding method will improve the robustness of the encoding through tougher consideration of the biochemical constraints, without affecting the compression performance of the original coder.
To ensure decodability, this decision has to be deterministic, otherwise, additional data would have to be transmitted to make sure the decoder can decode the data. 
\par
Overall, the proposed solution is very well adapted to block-based encoding methods like for example JPEG-inspired algorithms. 
More specifically in this work, codes are rotated after encoding a block. 
The order of encoding of the blocks is organized one line of blocks after another.
In addition, to avoid error propagation, the code decision process is reinitialised after the encoding of a full line of blocks as shown in Fig. \ref{fig:blocks}.

\subsection{Codes construction}

\label{subsec:construction}
The codes construction process is described in Algorithms \ref{algo:cdbks_gen} and \ref{algo:cw_mod}. The general idea is based on using an already existing codec and the associated quaternary code. From this pre-defined code, we will build three associated codes by 
mapping all the nucleotides in the original codewords to new ones according to a mapping rule
as described in Algorithm \ref{algo:cw_mod}. The results of the construction algorithm is a list of four codes that can be used in the novel solution. As proven by Proposition 1, the performance of the resulting coding solution is not affected by the introduction of the new coding method in comparison to the original coder.

\begin{algorithm}[h!]
\begin{algorithmic}
\caption{$GenerateCodes(InputCode)$: Create all the possible codes.}
\label{algo:cdbks_gen}
\STATE $Codes \leftarrow EmptyList(4)$
\STATE $Codes[0] \leftarrow InputCode$ 
\FOR{$i \leftarrow 2$ to $length(Codes)$}
\STATE $Codes[i] \leftarrow EmptyList(length(InputCode))$
\FOR{$j \leftarrow 1$ to $length(InputCode)$}
\STATE $Codes[i][j] = switchLetters(Code[j], i)$ 
\ENDFOR
\ENDFOR
\RETURN $Codes$
\end{algorithmic}
\end{algorithm}
\begin{algorithm}[h!]
\begin{algorithmic}
\caption{$switchLetters(codeword, k)$: Offset the letters of a codeword}
\label{algo:cw_mod}
\STATE $Mapping \leftarrow ['A', 'T', 'C', 'G']$
\FOR{$i \leftarrow 1$ to $length(codeword)$}
\STATE $idx \leftarrow index(codeword[i], mapping)$ 
\STATE $codeword[i] \leftarrow mapping[(i+k)$ mod $4]$
\ENDFOR
\RETURN $codeword$
\end{algorithmic}
\end{algorithm}
\noindent\textbf{Definitions.}
\textit{
\begin{itemize}
    \item S the source of length n
\vspace{-0.3\baselineskip}
    \item \(\Sigma\) the source S alphabet, \(S\in{\Sigma^n}\)
\vspace{-0.3\baselineskip}
    \item \(\Sigma_{DNA}\) the set \(\{A, T, C, G\}\) of nucleotides
\vspace{-0.3\baselineskip}
    \item \(C\) the original quaternary coder used to encode \(S\)
\vspace{-0.3\baselineskip}
    \item \(D=\{C_x\in{\Sigma_{DNA}^k}, \forall x\in{\Sigma}\}\) the code of \(C\)
\vspace{-0.3\baselineskip}
    \item \(C_R\) the novel quaternary coder
\vspace{-0.3\baselineskip}
    \item \(D_R=\{C_{Rx}\in{\Sigma_{DNA}^k}, \forall x\in{\Sigma}\}\) the code of \(C_R\)
\end{itemize}
}
\vspace{\baselineskip}
\noindent\textbf{Proposition 1.}\vspace{0.3\baselineskip}\\
\label{proposition_perf}
\textit{
\(Length(C(S)) = Length(CR(S))\), \vspace{0.3\baselineskip}\\
\hspace*{0.3cm}\(C\) and \(CR\) have the same compression rate.\vspace{0.5\baselineskip}\\
}
\noindent\textit{Proof}.\vspace{0.3\baselineskip}\\
\(Length(C(S)) = \sum_{i=1}^{Length(S)} Length(D[S[i]])\)\vspace{0.3\baselineskip}\\
\(Length(C_R(S)) = \sum_{i=1}^{Length(S)} Length(D_R[S[i]])\)\vspace{0.3\baselineskip}\\
Trivially, \(\forall x\in{\Sigma}, Length(D[x]))=Length(D_R[x])\)\\
\(\Rightarrow Length(C(S)) = Length(CR(S))\)

\subsection{Code choice}
\label{subsec:choice}
For the proposed coding solution, it is possible to choose between four codes to encode a fragment of the source. We propose two possible decision processes to determine the order in which the different codes will be used. The first possible case is very simple: we iterate through the four possible codes one by one, and start over. The second possible decision process is a pseudo-random choice, where the encoder and the decoder have a common seed for the pseudo-random generator. Introducing variability in the coding methods will result in a less structured quaternary stream, which is beneficial for our encoding process (less patterns and homopolymers, more balanced data for the GC content). 
\begin{figure}
    \centering
    \includegraphics[scale=0.3]{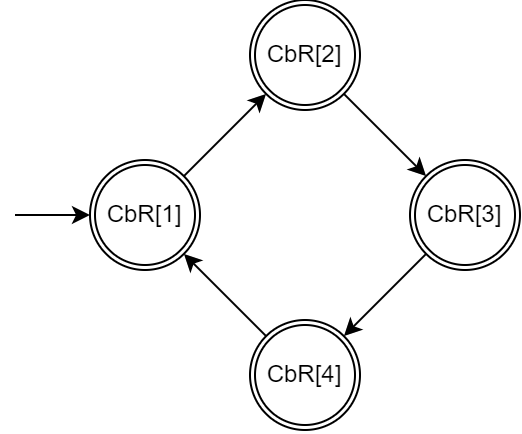}
    \caption{Codebook decision process, each node is one of the codebooks.}
    \label{fig:cdbk_automata}
\end{figure}
\begin{figure}
{\footnotesize
\begin{verbatim}
import random
def choose_random_codebook(codebook_candidates):
    seed(k)
    while True:
        i = randint(0, 3)
        yield codebook_candidates[i%4]
\end{verbatim}
}
    \caption{A python code for the pseudo-random codebook decision.}
    \label{fig:pseudo_random}
\end{figure}
\subsection{Integration in a block-based coder}
\begin{figure}
\centering
\begin{minipage}[c]{0.49\linewidth}
			\centerline{
              \includegraphics[scale=1]{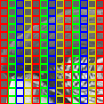}}
              \centerline{
              \footnotesize
		        \begin{tabular}{c}
                  (a) Simple rotation
	    	    \end{tabular}}
              \end{minipage}
 \begin{minipage}[c]{0.49\linewidth}
    		\centerline{
             \includegraphics[scale=1]{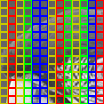}}
              \centerline{
              \footnotesize
		        \begin{tabular}{c}
                  (b) Pseudo-random rotation	
	    	    \end{tabular}}
\end{minipage}

\caption{Examples of rotation schemes on a block-based image coder. Here, the blocks colored in red are encoded with $C$, in green by $C^*_1$, in blue by $C^*_2$ and in yellow by $C^*_3$. 
}

\label{fig:blocks}

\end{figure}

In the case of block-based coders like the JPEG-inspired codecs, we can change the selected code every time we encode a new block. When the input of the block-based codec is an image, we can reduce the propagation of errors by reinitializing the decision process every time we start to encode a new line of blocks. 
Fig. \ref{fig:blocks} represents possible rotation schemes for a block-based image coding method. Every color represents the choice of a different code.

\subsection{About the fixed-length coders}
An important question to address about the proposed coder is why would we limit its usage only to variable-length coders. For example, the JPEG-inspired codecs don't exclusively use a variable-length coder, they also use a fixed-length coder. Ideally, using this method on any type of coder would improve the quality of the quaternary stream. But since a fixed-length coder is source independent, it can be pre-generated in a way to consider the constraints much more thoroughly. In such codes, specific codewords can be eliminated being considered as slightly problematic. If we rotated the letters similarly to variable-length coders, such problematic codewords would sometimes be created, losing all the quality gathered in this pre-generated code. 

%% file: performance_results.tex
\section{Performance results}
\label{sec:performances}
\subsection{Nomenclature}
For clarity, in the following sub-sections, the different methods will be named in the following logic:
\begin{center}
JPEGDNA-\$\{entropy coder id\}-\$\{source id\}[-\$\{rotation id\}]
\end{center}

The entropy coder id can be HG if the method uses the Huffman/Goldman coder or SFC4 if it uses the Shannon Fano Constrained coder. The source id can be S when the source is a raw image, or T when it is a already JPEG compressed binary file. For example, JPEGDNA-HG-S is the codec that encodes a raw image where the Huffman/Goldman coder encodes the run/category.
The rotation id is optionnal and can be either R or RR. The first describes a regular rotation scheme and the latter a pseudo-random rotation. Both rotations will be described in \ref{subsec:choice}. For example, the codec JPEG-DNA-HG-S-R is the equivalent of the latter but with a rotative labelling on the Huffman/Goldman coder.
\subsection{Compression performance}
\label{sec:CompressionPerformances}
\begin{figure}
\centering
\begin{minipage}[c]{0.49\linewidth}
			\centerline{
              \includegraphics[scale=0.47]{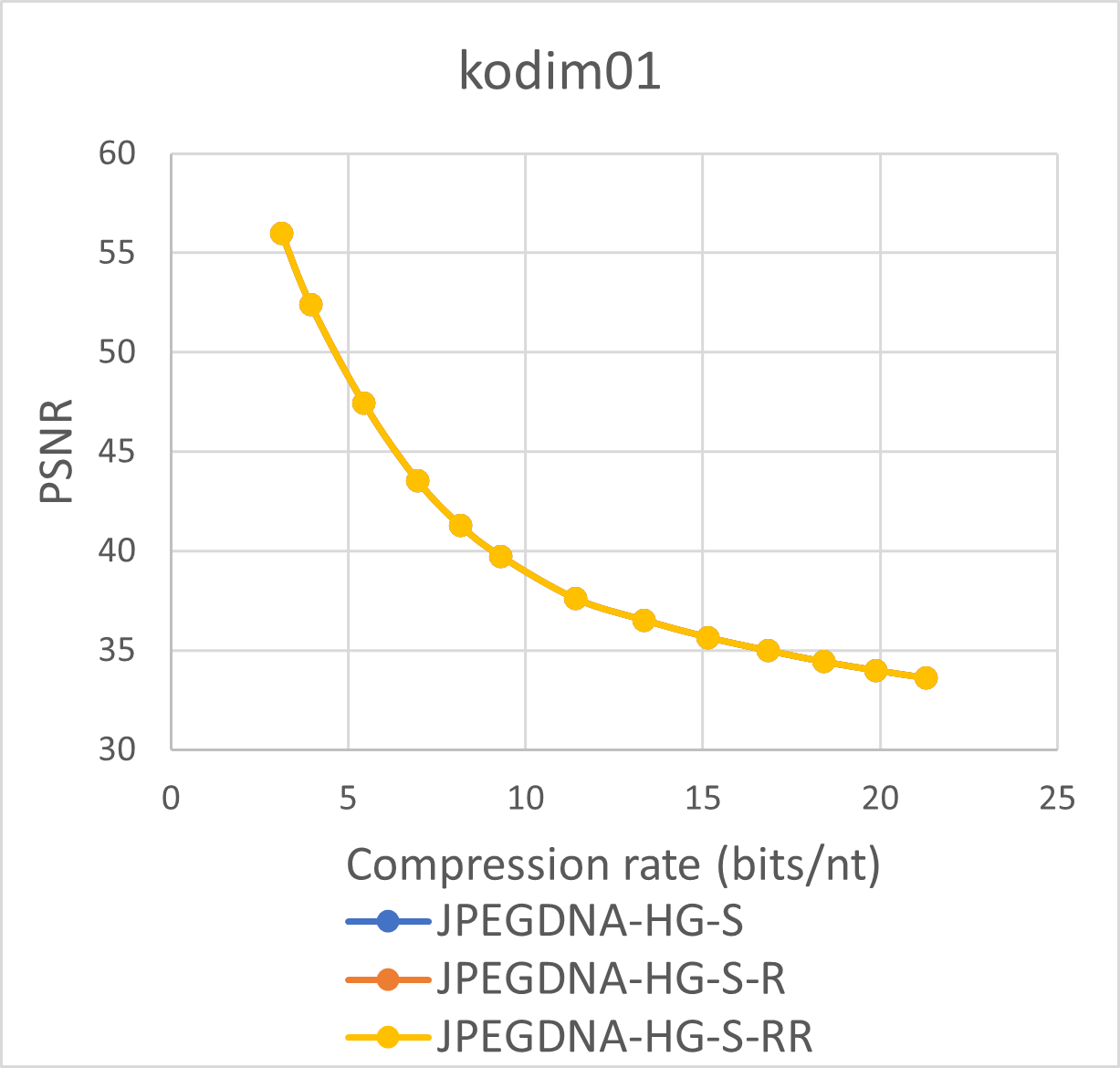}}
              \centerline{
              \footnotesize
		        \begin{tabular}{c}
                  (a) JPEG-HG-S
	    	    \end{tabular}}
              \end{minipage}
 \begin{minipage}[c]{0.49\linewidth}
    		\centerline{
             \includegraphics[scale=0.47]{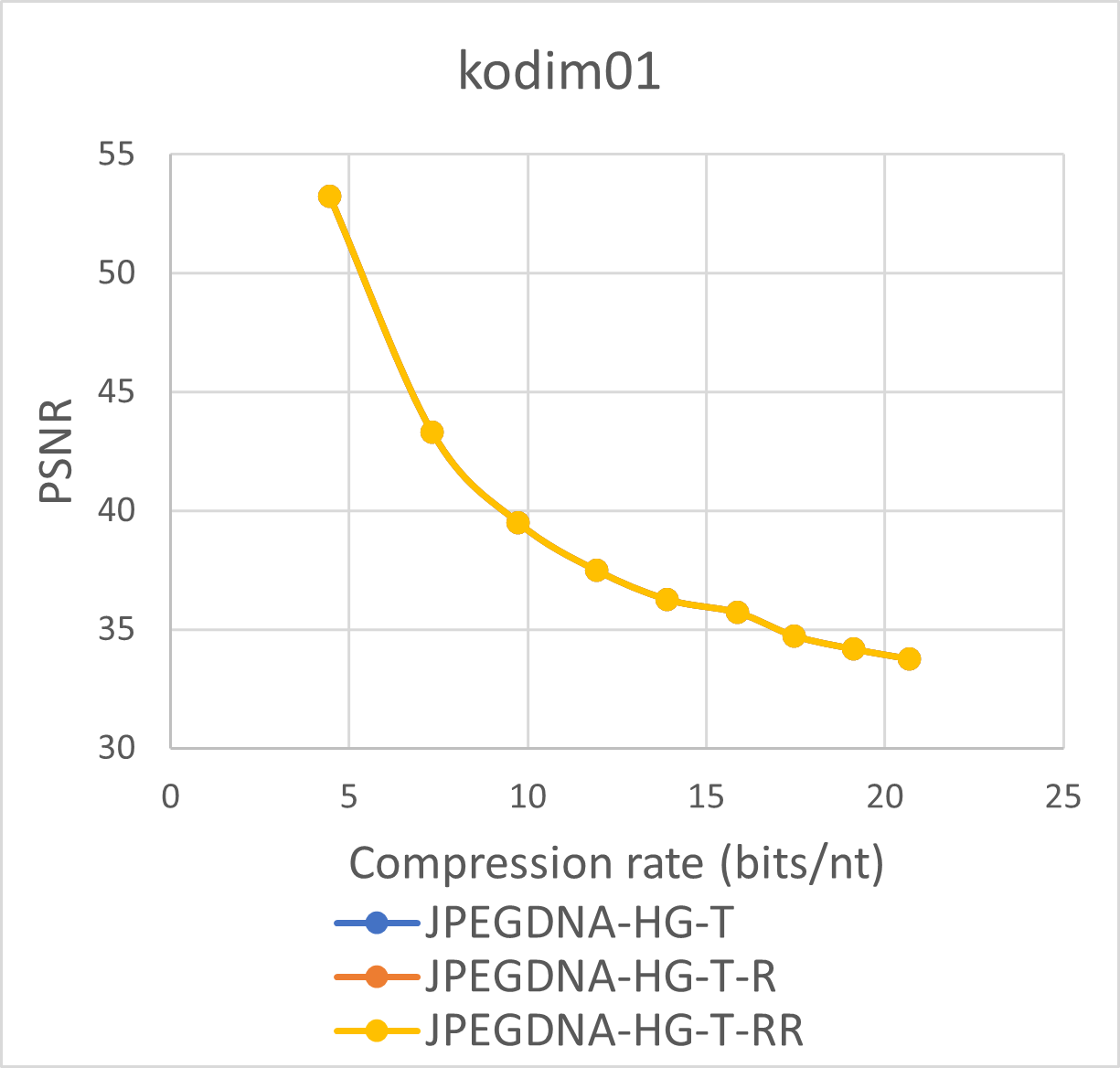}}
              \centerline{
              \footnotesize
		        \begin{tabular}{c}
                  (b) JPEG-HG-T
	    	    \end{tabular}}
\end{minipage}
\vspace{-1\baselineskip}
\caption{Performance comparison of the original algorithms and their novel modifications for two coding algorithms (JPEG-HG-S and JPEG-HG-T) on kodim01. The blue, yellow and red curves are superimposed.}
\label{fig:rate-dist}
\end{figure}

As previously shown in subsection \ref{subsec:construction}, the introduction of the new coding method does not have any effect on the coding performance in comparison to the original method. However, this property has only been proven specifically for the rotating codes, 
and not for the codecs in which this new method is integrated. But since we only modify the variable-length coder and not the rest of the compressed stream in those JPEG-inspired coding methods, the compression rate of the general coding algorithm remains the same. The rate-distorsion curves in Fig. \ref{fig:rate-dist} illustrate this: the curves for the original algorithms and the modified ones are superimposed. 
The performance results shown in Fig. \ref{fig:rate-dist} have been computed on the first image of the Kodak dataset\footnote{\color{blue}\url{https://r0k.us/graphics/kodak/}}, \textit{kodim01}.

%% file: stats_review.tex
\subsection{Oligo quality assessment}
\label{sec:quality}
\begin{figure*}
    \centering
    \includegraphics[scale=0.65]{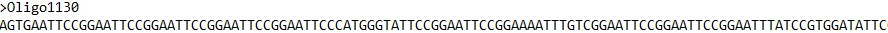}
    \vspace{-\baselineskip}
    \caption{Result after rotating the variable-length code. This oligo and the one in Fig. \ref{fig:JPEGDNAprob} represent exactly the same data, and in comparison to Fig. \ref{fig:JPEGDNAprob}, homopolymers have disappeared. 
    }
    \label{fig:JPEGDNARprob}
    \vspace{-1.5\baselineskip}
\end{figure*}
\subsubsection{Approach}
Respecting the biochemical constraints of DNA data storage is crucial for the end-to-end storage. 
This is reflected in Fig. \ref{fig:JPEGDNAprob}, where, due to the apparition of very long homopolymers, the obtained oligos are unusable. In comparison, the same oligos obtained for the same compression rate with the  modified method of the algorithm, shown in Fig. \ref{fig:JPEGDNARprob} are much more acceptable from the biochemical perspective. Since a visual verification is not enough to assess the quality of the generated oligos, we further introduce some oligo quality analysis and visualization tools that were, to our knowledge, lacking in the field of DNA data storage. 
\subsubsection{Software for quality assessment}
The proposed analysis and visualization tools are available in a public repository\footnote{\color{blue}\url{https://github.com/jpegdna-mediacoding/OligoAnalyzer}}. 
These tools inlude general statistics about the coded data (number of homopolymer in the oligo pool, average size of the homopolymers, average GC content, proportion of oligos with problematic GC content) as well as histograms describing with more details the ill cases. To our knowledge, in past research \cite{DNASmart} , gathering statistics to measure the quality of oligos have never been presented in depth, especially for the case of homopolymers.
\subsubsection{GC content}
\begin{figure}
    \centering
    \includegraphics[scale=0.45]{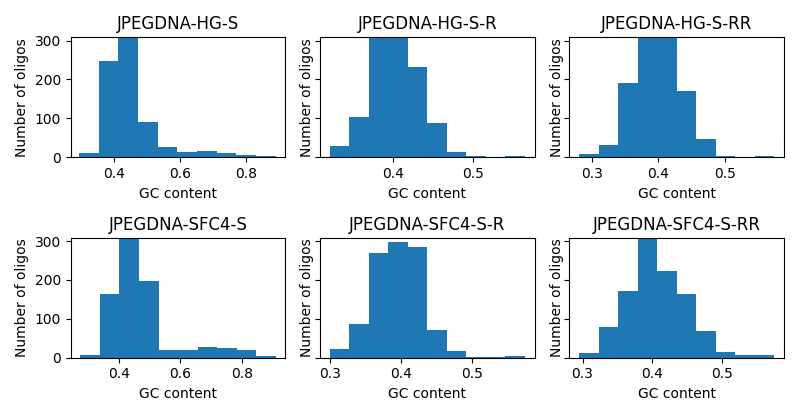}
    \vspace{-2\baselineskip}
    \caption{Distribution of the generated oligos when encoding kodim01 at 30 bits/nt in function of their GC content. The novel methods don't have problematic oligos (above 60\% of GC).}
    \label{fig:gc_content}
    \vspace{-1.5\baselineskip}
\end{figure}

The distribution of the GC content of the oligos has been evaluated. We consider all the oligos with a GC content below 30\% or above 60\% to be problematic.
In these regards, as presented in Fig. \ref{fig:gc_content}, the novel methods (JPEGDNA-HG-S-R, JPEGDNA-HG-S-RR, JPEGDNA-HG-T-R, JPEGDNA-HG-T-RR) show great improvement over the original ones.
\subsubsection{Homopolymers}
\begin{figure}
    \centering
    \includegraphics[scale=0.50]{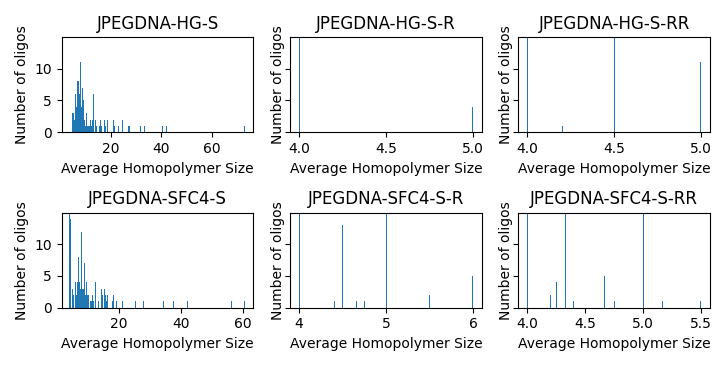}
    \vspace{-2\baselineskip}
    \caption{Distribition of the average homopolymer size per oligo when encoding kodim01 at 30 bits/nt. The novel methods don't have oligos with very long homopolymers.}
    \label{fig:homopolymers_avg_size}
    \vspace{-1.5\baselineskip}
\end{figure}

For every oligo, we have computed the average length of homopolymer runs. At high compression rates, the homopolymer length deteriorates quickly and some oligos have homopolymers of a size greater than ten, which is extremely problematic, as shown in Fig. \ref{fig:homopolymers_avg_size}. Even though the novel methods still produce some homopolymers, their average size per oligo does not exceed the length of five nucleotides, which is a very important gain for the robustness of the decoded result. Moreover, the number of homopolymers and their average size over the whole oligo pool also decreased as shown in Table \ref{tab:homopolymers}, where $N$ is the number of homopolymers in the data, $Avg$ the average size of the homopolymers and $Max$ the maximum size of the homopolymers. The pseudo-random rotation slightly underperforms in terms of length of homopolymer runs in the compressed data. This is due to the fact that in some cases, the code decision process will pick the same code several times in a row, allowing the apparition of more homopolymers. The SFC4-based \cite{SFC4} coders also underperform, since they were designed to relax the homopolymer constraint in comparison to Huffman/Goldman.
\begin{table}[h]
    \caption{Homopolymers in the oligos when encoding kodim01 at  30 bits/nt
    }
    \centering
    \begin{tabular}{|c|c|c|c|c|c|c|}
        \hline
        & HG-S & HG-S-R & HG-S-RR & SFC4-S & SFC4-S-R & SFC4-S-RR\\
        \hline
        N & 756 & \textbf{64} & 291 & 939 & \textbf{411} & 644 \\
        Avg & 10.68 & \textbf{4.11} & \textbf{4.14} & 9.38 & \textbf{4.18} & \textbf{4.20} \\
        Max & 67 & \textbf{5} & \textbf{5} & 57 & 6 & 7 \\
        \hline
    \end{tabular}
    \label{tab:homopolymers}
\end{table}

%% file: conclusion.tex
\section{Conclusion}
In this paper, we have proposed a novel coding method adapted to DNA data storage that introduces variability in the generated quaternary streams. 
The main asset of the proposed encoding algorithm lies in its ability to tackle some of the challenges that are met when designing coders for DNA data storage, namely the respect of biochemical constraints. Results show that the generated oligos contain less homopolymer runs while the remaining ones present a greatly reduced size. 
Moreover, results show improvements on the GC content, limiting it between 30\% and 60\%.
The improvements were more significant at lower bit-rates, where the original coding algorithms severely underperformed with regards to the biochemical constraints.
Additionally, thanks to its design, integrating the proposed coding solution in block-based image codecs does not
affect the compression rate of the original solution.